\documentclass[a4paper]{article}
\usepackage{smsi}
\usepackage{graphicx}
\usepackage{amsmath}
\usepackage{siunitx}
\usepackage[hidelinks]{hyperref}   
\usepackage{orcidlink}             
\newcommand{\doi}[1]{doi: \href{https://doi.org/#1}{#1}}
\graphicspath{{figures/}}
\begin{document}

\title{Continuous Phase-Coherent Readout of a Magneto-Mechanical Resonator Using a Phase-Locked Loop}

\author{\underline{Frauke H. Niebel}\textsuperscript{1}\,\orcidlink{0000-0003-3394-0122},
        Jan-Philipp Scheel\textsuperscript{1,2}\,\orcidlink{0000-0002-7840-4879},
        David Conrad\textsuperscript{1}\,\orcidlink{0009-0009-2156-9419},
        Hossein Shabanalinezhad\textsuperscript{1}\,\orcidlink{0009-0007-2693-8899}\\  
        \itshape and Matthias Graeser\textsuperscript{1}\,\orcidlink{0000-0003-1472-5988}
        }

\renewcommand{\institute}{\textsuperscript{1} Metrology Lab, Faculty of Computer Science and Electrical Engineering, University of Rostock, Rostock, Germany,
          \{frauke.niebel,matthias.graeser\}@uni-rostock.de \\
          \textsuperscript{2} Fraunhofer Research Institution for Individualized Medical Technology and Engineering IMTE, L{\"u}beck, Germany \\
          }

\renewcommand{\email}{} 

\renewcommand{\abstractt}{
Magneto-mechanical resonators are wireless sensors that encode a measurand in their natural
frequency. Existing systems estimate this frequency from repeated cycles of excitation and
ring-down, which limits the precision and the bandwidth range. Here, we introduce a continuous,
phase-coherent readout based on a phase-locked loop. In a benchtop setup, one record stayed within
4~\% of the precision--bandwidth limit over more than three decades of bandwidth, while conventional
readouts remained about a factor of 4 above it.}

\renewcommand{\keywords}{magneto-mechanical resonator, phase-locked loop, frequency estimation, wireless sensing, readout}

\maketitle

\section{Background, Motivation and Objective}
Magneto-mechanical resonators (MMRs) are an emerging class of passive, wireless sensors that exploit
the mechanical resonance of a permanent magnet~\cite{bib:gleich}. They retain a strong signal at
volumes below 1~mm$^3$, which makes them candidates for long-term monitoring of physiological
parameters~\cite{bib:gleich}. The measurand is encoded in the natural frequency of the MMR's
torsional oscillation, which must be estimated from a voltage signal. As any uncertainty in this
estimate directly propagates to the measured parameter, frequency estimation is critical to MMR
sensing.

Frequency estimation from noisy samples is bounded by a trade-off between precision and bandwidth.
Thus, a readout is judged by two criteria: its precision, evaluated by how closely it approaches the
theoretical limit, and its bandwidth range. The latter matters as physiological measurands often
span multiple timescales, from slow trends to fast waveforms~\cite{bib:abraham}.

State-of-the-art readouts observe the resonator in consecutive frames~\cite{bib:gleich,bib:reiss}.
Each frame excites the resonator, records its free decay, and estimates the frequency anew. This
gated architecture keeps the drive from coupling directly into the receiver but incurs four
penalties: (i) the receiver is blind during excitation, (ii) the undriven signal decays during
reception, so the signal-to-noise ratio decreases toward the end of each receive window, (iii) each
frame discards the phase history of the previous one, so lower bandwidths can only average
independent frames, and (iv) the frequency is estimated once per frame, so variations above half the
frame rate alias. Together, these penalties reduce the attainable precision and bandwidth range.

\begin{figure}[t]
  \centering
  \includegraphics[width=\columnwidth]{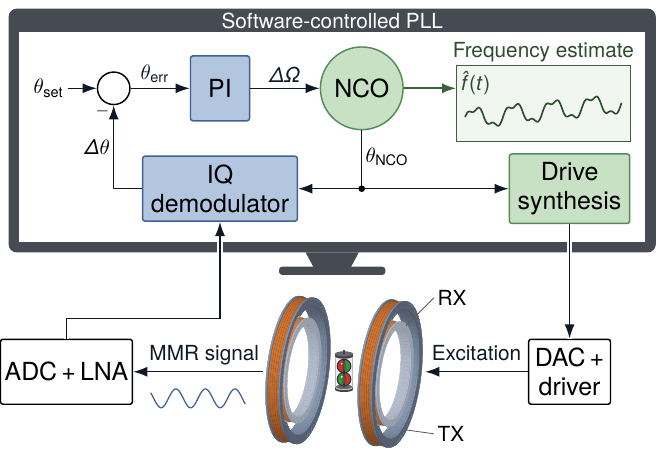}  
  \caption{PLL readout architecture. The PI controller corrects the NCO frequency from the IQ phase
  error. This frequency is the estimate. Feedthrough compensation coils are not shown.}
  \label{fig:concept}
\end{figure}

Here, we present an MMR readout that avoids these penalties. It uses a phase-locked loop (PLL), a
well-established technique in atomic force microscopy~\cite{bib:giessibl}. The PLL drives the MMR
continuously and tracks its phase coherently. We demonstrate it on a benchtop MMR setup and compare
it to gated readouts.

\section{Methods and Experiments}
Figure~\ref{fig:concept} shows the PLL readout architecture. A numerically controlled oscillator
(NCO) generates the excitation reference, whose phase is compared with that of the received signal
through in-phase and quadrature (IQ) demodulation. A proportional--integral (PI) controller turns
the phase error into a correction of the NCO frequency. This keeps the resonator synchronized with
the drive while the adjusted NCO frequency serves as the frequency estimate. Due to the MMR's
nonlinear dynamics, this frequency deviates from the natural frequency by a
deflection-angle-dependent shift~\cite{bib:gleich}. In this work, we consider a fixed MMR pose and assume
the measurand alone determines the deflection angle, so a calibration can largely absorb this shift.

The PLL runs in software on a benchtop setup. An NI USB-6453 data acquisition device (DAQ) outputs
the drive signal to a current-controlled driver. The driver excites the MMR (natural frequency
$f_0\approx318$~Hz, time constant $\tau\approx1.8$~s, quality factor $Q\approx1800$), which is fixed
at the center of a Helmholtz coil pair. A concentric second pair receives the MMR signal, which is
amplified and digitized. As the PLL drives and receives simultaneously at the same frequency, the
drive couples directly into the receive coils and biases the measured phase. To suppress this drive
feedthrough, an identical transmit and receive set without an MMR is connected in opposite polarity.

The theoretical precision--bandwidth limit serves as the benchmark. For white receiver noise, the
smallest frequency noise any readout can achieve at a given oscillation amplitude when resolving
measurand variations up to a bandwidth $B$ is
\begin{equation}
  \sigma_f^2\approx\frac{N_0}{C}\Big(\frac{B^3}{3}+\nu_\mathrm{L}^2B\Big),\quad \nu_\mathrm{L}=\frac{1}{2\pi\tau},
\end{equation}
where $C/N_0$ is the carrier-to-noise density ratio. The laboratory sensing system contributes
unrealistically little noise, so that slow fluctuations and drift of the sensor dominate the
frequency noise, masking differences between readouts. To mimic the lower $C/N_0$ of a small, more
distant MMR in a clinical-scale system, we set $C/N_0$ to 40~dB-Hz. Given this level and $\tau$
measured from a free decay, we computed the limit over the band from 0.01~Hz to $B$.

We derived the PLL's $\sigma_f(B)$ from a 30-minute record at a drive amplitude of 3~µT, which
sustained a deflection angle of about 9.5$^\circ$. To reach 40~dB-Hz, we added white Gaussian noise
to each sample in the running loop. Near and above the loop bandwidth $B_\mathrm{L}$ of 10~Hz, the
loop response distorts this record, which we corrected offline using the logged phase error, the
loop delay and the resonator's phase dynamics. A Bode measurement of known $f_0$ modulation showed
that the corrected estimate reproduces measurand variations to within 4~\% up to 100~Hz. Integrating
the one-sided power spectral density of the corrected frequency noise from 0.01~Hz to $B$ gives
$\sigma_f^2(B)$.

For reference, we implemented two gated readers on the same setup. A free-running NCO drove the MMR
in bursts of 7.8~µT, resulting in the same time-averaged drive as for the PLL. Following Gleich et
al.~\cite{bib:gleich}, the receive window $T_\mathrm{RX}$ takes about 60~\% of each frame. We
considered two frame rates $f_\mathrm{r}$: a fast reader using their published timing
($T_\mathrm{RX}=49$~ms, $f_\mathrm{r}=12.2$~Hz) and a slow reader scaling this timing to the time
constant $\tau$ of our MMR ($T_\mathrm{RX}=1.13$~s, $f_\mathrm{r}=0.54$~Hz). A maximum-likelihood
fit of a decaying sinusoid to each receive window gave one frequency estimate per frame. We added
noise of the same density offline and processed the estimates as for the PLL, up to
$f_\mathrm{r}/2$.

\begin{figure}[t]
  \centering
  \includegraphics[width=\linewidth]{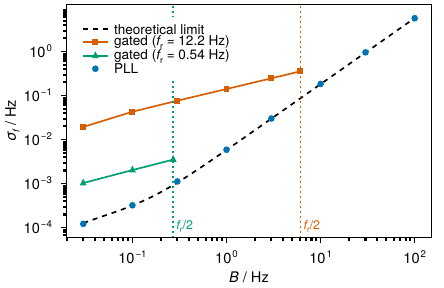}
  \caption{Standard deviation $\sigma_f$ of the frequency estimate versus measurand bandwidth $B$ at
  $C/N_0=40$~dB-Hz for the PLL and two gated readers. Dashed: theoretical limit.
  Dotted: $f_\mathrm{r}/2$.}
  \label{fig:sigmaf}
\end{figure}

\section{Results and Discussion}
In Fig.~\ref{fig:sigmaf}, the PLL stays within 4~\% of the theoretical limit from 0.03~Hz to 100~Hz.
The gated curves end at $f_\mathrm{r}/2$ (6.1~Hz and 0.27~Hz) and stay above the limit because of
two limitations.

The first limitation is a narrow bandwidth range. The frame rate sets both the sampling rate of the
estimates and the duration of each receive window. Through aliasing (iv), the sampling rate caps the
bandwidth at $f_\mathrm{r}/2$. Through the loss of phase history (iii), the window duration sets the
precision at all lower bandwidths, as frames can only be averaged incoherently. This coupling
imposes competing demands: reach calls for high frame rates, precision for low ones. The optimal
frame rate is the lowest that avoids aliasing, $f_\mathrm{r}=2B$, with two implications: the target
bandwidth must be known before acquisition, and a record is optimized for only one bandwidth.

The second limitation is reduced precision. Even at $f_\mathrm{r}/2$, the gated readers stay above
the limit. The gap can be attributed to penalties (i)--(iv), which prevent efficient use of the
available resonator signal. Dead time (i) and the one-estimate-per-frame structure (iii, iv) account
for a gap of about a factor of 4 at $f_\mathrm{r}/2$. Signal decay (ii) barely affects the fast
reader but limits the benefit of longer windows at the slow end.

The PLL avoids both limitations by continuously observing the resonator and keeping its phase, at
the cost of handling drive feedthrough. The bandwidth can thus be chosen in the analysis while
staying close to the theoretical limit.

\end{document}